\documentclass[]{revtex4}
\usepackage{bm}
\usepackage{lettrine}
\usepackage{dcolumn}
\usepackage{bm}
\usepackage{graphicx}
\usepackage{amsmath}
\usepackage{latexsym}
\usepackage{amsfonts}
\usepackage{amssymb}
\usepackage{array}
\usepackage{epsfig}
\usepackage{epstopdf}
\usepackage{times}
\usepackage{amsmath,epsfig}

\usepackage{tikz}

\newcommand{\circled}[1]{\tikz[baseline=(char.base)]{\node[shape=circle,draw,inner sep=0.7pt] (char) {#1};}}
\newcommand{\ket}[1]{\left\vert#1\right\rangle}
\newcommand{\bra}[1]{\left\langle#1\right\vert}

\begin{document}

\title{A no-go result on the purification of quantum states}

\author{Carlo Di Franco and Mauro Paternostro}
\affiliation{Centre for Theoretical Atomic, Molecular and Optical Physics, School of Mathematics and Physics, Queen's University, Belfast BT7 1NN, United Kingdom}

\begin{abstract}
{\bf The information encoded in a quantum system is generally spoiled by the influences of its environment, leading to a transition from pure to mixed states. Reducing the mixedness of a state is a fundamental step in the quest for a feasible implementation of quantum technologies. Here we show that it is impossible to ``transfer'' part of such mixedness to a ``trash'' system without losing some of the initial information. Such loss is lower-bounded by a value determined by the properties of the initial state to purify. We discuss this interesting phenomenon and its consequences for general quantum information theory, linking it to the information theoretical primitive embodied by the quantum state-merging protocol and to the behaviour of general quantum correlations.}
\end{abstract}

\maketitle

\lettrine{I}n the {\it ``whisper-down-the-lane''} game, players are aligned to form a chain and the first player whispers a message to his nearest neighbour. Each player then does the same with the next one down the chain, until the message reaches the last person. Clearly, errors typically accumulate in the process (each player passes what he {\it believes} is the message to share)  so that the statement that is revealed to the last in the lane may be significantly different from the original one: the noise affecting the information travelling along the lane has spoiled its quality. Needless to say, a reliable communication channel (of which the above is definitely not an example!) should be such that the input and output messages overlap quite significantly, if not perfectly, regardless of the message, its complexity and the actual length of the channel itself. In order to counteract the degradation of the message's quality, a classical communication channel is often interspersed by amplifiers and filters, aiming at increasing the signal-to-noise ratio and thus getting a better quality transmission.

\begin{figure}[t]
\psfig{figure=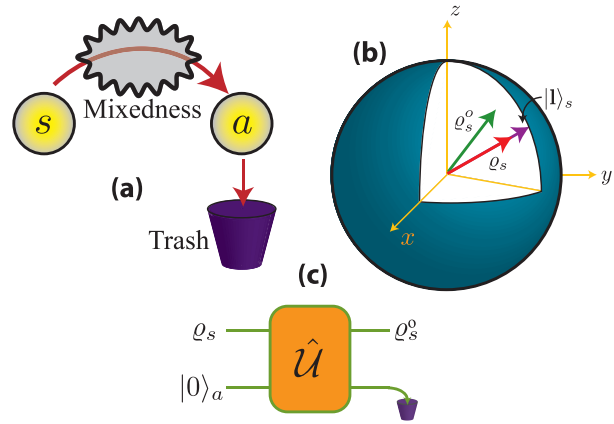,width=8cm}
\caption{{\bf (a)} Sketch of the investigated scenario: part of the mixedness present in system $s$ is transferred to the ancillary system $a$, which is discarded after the operation. {\bf (b)} Bloch-sphere picture of the process. We show the vectors representing the mixed initial state $\varrho_s$ that should be purified, the pure reference state $\ket{\bf l}_s$ and the output state $\varrho_s^o$ resulting from the application of our protocol. A successful protocol requires a small angle between the vectors representing $\varrho_s$ and $\varrho^o_s$ and the latter being longer than the former. {\bf (c)} The whole scheme can be implemented on a quantum circuit exploiting a joint unitary transformation $\hat{\cal U}$ over the unknown state $\varrho_s$ of the system and state $\ket{0}_a$ of the ancilla.}
\label{scheme}
\end{figure}

Quantum mechanically, it is often the case that the message to transmit is the pure quantum state of a system. Such state will be acted upon by the surrounding world, during its transmission, through decoherence mechanisms that reduce its purity~\cite{nielsenchuang}. The latter is an indicator of the knowledge that we have on the preparation of the system: losing purity (or, equivalently, making the state more mixed), implies pushing the system more ad more towards classicality, losing at the same time information on the original message itself. A fully mixed state is just a classical uniform probability distribution to find the system in one of its possible physical configurations, with no quantum feature left. In this respect, purification may be the key~\cite{cirac}: by acting on the output state, using many copies of it, the interactions with some ancillae and measurements, one can indeed retrieve part of the information lost during the communication process. Experimentally, state purification has been demonstrated in linear optics for the case of two copies of the state to manage~\cite{ricci}.

However, how close would the purified state be with respect to the original message that we aimed at sending off? This question introduces another form of {\it faithfulness} that we should take care of, intuitively related to the quality of the message transmitted across the lane of whisperers above. Moreover, it is very interesting to determine whether a {\it global} improvement is possible, where the quality of the exchanged message is high with respect to chosen purity and closeness indicators. Can we achieve an output state arbitrarily close to the input and, at the same time, gain purity at the expenses of the noise? Classically, nothing seems to prevent this.

Here we show that, quantum mechanically, this is certainly not the case: by designing a special ancilla-based purification protocol, we find the existence of  a trade-off that prevents the purification of a state that also remains close at will to the input one. Such an impossibility is strongly related to fundamental features of the system-ancilla state, whose nature as a resource for quantum communication goals determines the efficiency of the global optimization task mentioned above. See Refs.~\cite{kleinmann} for studies closely related to our goals. 

The scheme we consider is based on the idea of removing part of the mixedness present in our state without losing any of the information encoded in it, or at least trying to minimize this loss. We do this using the ancilla as a a ``trash'', which is discarded at the end of the process, as sketched in Fig.~\ref{scheme} {\bf (a)}. To address our question without unnecessary complications, we consider our input message as encoded in the state of a two-level quantum system (labeled as $s$). The ``trash'' ancilla (labeled as $a$) will also be two-dimensional, while the model can clearly be extended to higher-dimensional cases.

\begin{figure}[b]
\psfig{figure=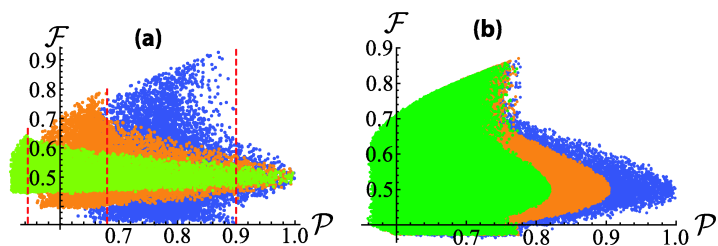,width=9.5cm}
\caption{{\bf (a)} Trade-off between the information gathered on the state of $s$ and the final purity achieved through the protocol. Both the fidelity ${\cal F}$ and purity ${\cal P}$ are the results of an average over randomly generated initial states of $s$ with a set value of purity ${\cal P}_{in}$. We have taken the ancilla as prepared in $\ket{0}_a$ and ${\cal P}_{in}=0.545,0.68$ and $0.905$ [corresponding to $p_w=0.3,0.6$ and $0.9$ in Eq.~\eqref{state}], going from the light-colored to the dark-colored points respectively (green, orange, blue). The dashed vertical lines show the points at which the output purity equals the input one. {\bf (b)} Same as panel {\bf (a)} but for an ancilla prepared in a mixed state of purity $0.82,0.905$ and $1$ (from light-colored to dark-colored points) and ${\cal P}_{in}=0.78125$ ($p_w=0.75$).}
\label{distribuzioni}
\end{figure}

\begin{figure*}[t]
\psfig{figure=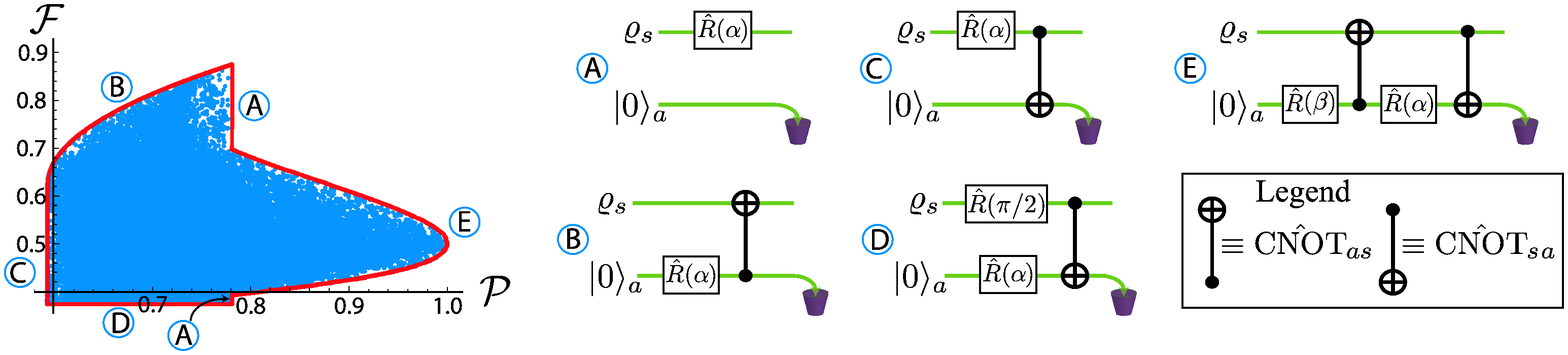,width=16cm}
\caption{Comparison between the distribution of output fidelity ${\cal F}$ and purity ${\cal P}$ for $N{=}10^5$ random unitaries (dark blue points) and the results achieved through numerical extremization (boundary red traits). We have taken $p_w=0.75$, corresponding to ${\cal P}_{in}=0.78125$ and ${\cal F}_{in}=0.875$. Each point is obtained as the ensemble average over $5\times10^4$ input states (all with the same initial purity ${\cal P}_{in}$). The boundary points result from the use of the quantum circuits identified by the circled letters shown in the figure.}
\label{gates}
\end{figure*}

As a measure of mixedness, we will consider the purity of the system under investigation defined as ${\cal P}{=}{\rm Tr}_s[\varrho^2_{s}]$, where ${\rm Tr}_i$ denotes the partial trace with respect to system $i{=}s,a$, $\varrho_s{=}{\rm Tr}_a[\varrho_{sa}]$ ($\varrho_a{=}{\rm Tr}_s[\varrho_{sa}]$) is the reduced density matrix corresponding to system $s$ (the ancilla $a$) and $\varrho_{sa}$ is the density matrix describing the quantum state of the joint system. We use the Bloch vector formalism~\cite{nielsenchuang} according to which a single-qubit state is in one-to-one correspondence with a vector ${\bf l}=(l_1,l_2,l_3)$. Pure (Mixed) states correspond to vectors of length equal to (smaller than) $1$ and thus lie on the surface (occupy the interior) of the so-call Bloch ball. The purity of a given state is related to the length $\ell$ of the vector ${\bf l}$ through the simple expression ${\cal P}=(1+\ell^2)/2$.

The other information benchmark that will be used throughout this work is precisely the direction of the Bloch vector. In the three-dimensional geometric space introduced above, two states with comparable purities are similar to each other if the corresponding Bloch vectors point along close directions. Therefore, given a state $\varrho_s$ to purify, we identify the direction of the associated vector ${\bf l}_s$ and consider the {``reference''} given by the {\it pure} state $\ket{\bf l}_s$, whose Bloch vector points precisely along ${\bf l}_s$. At the end of the purification protocol, we will distinguish between the length of the Bloch vector of the output state $\varrho^o_s$ ({\it i.e.} we will quantify its purity) and its direction with respect to the reference state $\ket{\bf l}_s$. The latter figure of merit is formally embodied by the so-called state fidelity ${\cal F}{=}{}_s\!\bra{\bf l}\varrho^{o}_{s}\ket{\bf l}_s$~\cite{nielsenchuang}, which is linearly dependent on the projection of the Bloch vector of the output state onto the vector associated with $\ket{\bf l}$ [the information initially encoded in the state of $s$ corresponds to ${\cal F}{=}(1+\ell)/2$]. The above discussion has thus sketched the situation that we consider in this work, as shown in Fig.~\ref{scheme} {\bf (b,c)}: given a state to purify, we device a protocol based on a joint unitary interaction $\hat{\cal U}$ with the ancilla $a$ (to be discarded later) such that the length of the Bloch vector of $\varrho^o_s$ has grown (which corresponds to an increased output purity) without changing its direction significantly (so that the fidelity with the reference state remains close to the initial value). In order to fix the ideas, here we consider initial states of $s$ having the form
\begin{equation}
\label{state}
\varrho_{s}=p_w\ket{\bf l}_s\!\bra{\bf l}+(1-p_w)\hat{I}/2
\end{equation}
with $\hat{I}$ the $2\times2$ identity matrix and $p_w{\in}[0,1]$. A simple calculation shows that $\ell{\equiv}p_w$, so that ${\cal P}_{in}{=}(1+p^2_w)/2$. Physically, Eq.~\eqref{state} corresponds to the action of a form of noise (sometimes referred to as {\it white noise}~\cite{nielsenchuang}) that shrinks ${\bf l}_s$ isotropically  from unity to its length $p_w$. However, this does not imply that we are restricting the study only to a specific form of noise: any single-qubit mixed state can be written as in Eq.~\eqref{state}.

The protocol should be state-independent, which means that, besides assigning the value of $p_w$ in Eq.~\eqref{state}, we do not impose any restriction to the form of $\ket{\bf l}_s$. Our figures of merit will thus be averaged over any possible choice of such pure-state component. On the other hand, we decide to prepare system $a$ in a fiducial state that, unless otherwise specified, we have taken to be $\ket{0}_a$. While any other pure-state preparation is equally legitimate, the assumption of pure ancillary state is important. It implies that the ``trash'' system $a$ is able to {\it accept} more of the mixedness that we aim at transferring (needless to say, due to the unitarity of $\hat{\cal U}$, the global mixedness of $\varrho_{sa}$ is preserved). A generalization of our results to initially mixed ``trash'' systems is also briefly discussed later on. As a final remark, we point out that we set no constraint on the form of the unitary interaction between $s$ and $a$ (which could also include single-qubit operations).
\\
\\
\noindent
{\large {\bf Results}}

\noindent
We have constructed $\hat{\cal U}$ by resorting to the theory of random unitaries~\cite{kus}. Starting from an ensemble of $M$ random mixed states having the form given in Eq.~\eqref{state} (such ensemble is used to calculate average values of the figures of merit under scrutiny here), we have applied $N$ random unitary operations (constructed using 15 parameters, uniformly drawn with respect to the proper Haar measure as described in {\bf Methods}) so as to obtain $N$ distributions of $M$ output states. The ensemble-averaged purities ${\cal P}$ and fidelities ${\cal F}$ with respect to the pure reference state have been calculated and plotted against each other for each of the $N$ random gates. Intuitively, one should expect the existence of a price to pay for transferring mixedness from $s$ to $a$ and that a successful purification protocol would necessarily deplete the information content as quantified by state fidelity. Such a prediction is indeed confirmed by the results shown in Fig.~\ref{distribuzioni} {\bf (a)}, where we see that, by chosing the proper set of unitaries, an arbitrary high degree of average purity is achievable through our protocol and regardless of the initial purity ${\cal P}_{in}$, although the fidelity with the reference state is strongly reduced. Moreover, as expected and discussed previously, the purity of the auxiliary qubit $a$ plays a key role in the performance of the protocol: by taking a mixed state of $a$ we limit its capability of receiving mixedness and so lower the efficiency of the purification protocol. Fig.~\ref{distribuzioni} {\bf (b)} shows that the range of average output purities achievable strongly depends on the mixedness of the ancilla. However, some undoubtedly striking features emerge from our random-unitary analysis, in particular with respect to the trade-off between the purification capabilities of the protocol and the output state fidelity.

Let us analyze these results more closely. The points along the vertical dashed lines in Fig.~\ref{distribuzioni} {\bf (a)} correspond to output purity equal to the input one. The top-most point has output state fidelity identical to the initial one and vertical trait is spanned by the cases corresponding to the application of local unitary operations on qubit $s$ (which do not alter the purity, yet reduce the fidelity). What is striking, though, is that as soon as we try to increase purity, {\it i.e.} we {move} to the right of the vertical line, we observe a large drop of fidelity: there is non-trivial trade-off between purity and fidelity. We are thus in a position to state the key point of our investigation, which can be formulated as the following {\it no-go result}:

\noindent
{\bf Statement}: {\it Under a mixedness-trashing protocol, purity cannot be increased past its initial value without reducing significantly the information content of the output state}.

In order to characterize properly the boundaries of our plots, we have performed a constrained-optimization with the following strategy. A heuristic analysis based on a numerical exploration suggests that the unitary gate $\hat{\cal U}$ in our general scheme can be easily decomposed in terms of only three building blocks~\cite{gatenote}. In the ordered two-qubit computational basis $\{\ket{00},\ket{01},\ket{10},\ket{11}\}_{as}$, these are given by the single-qubit rotations $\hat{R}(\alpha)=\cos\alpha\hat{I}+i\sin\alpha\hat\sigma_y$ and the two-qubit gates
\begin{equation}
\hat{\text{CNOT}}_{as}{=}
\left(
\begin{matrix}
1&0&0&0\\
0&1&0&0\\
0&0&0&1\\
0&0&1&0
\end{matrix}
\right),~~\hat{\text{CNOT}}_{sa}{=}
\left(
\begin{matrix}
1&0&0&0\\
0&0&0&1\\
0&0&1&0\\
0&1&0&0
\end{matrix}\right).
\end{equation}
$\hat{\text{CNOT}}_{as}$ is a {\it controlled-NOT} operation that flips (leaves unchanged) the state of $s$ when $a$ is prepared in $\ket{1}_a$ ($\ket{0}_a$). $\hat{\text{CNOT}}_{sa}$ flips (leaves unchanged) the state of $a$ when $s$ is prepared in $\ket{1}_s$ ($\ket{0}_s$). Needless to say, not all such gates are necessary in order to span the boundaries of the physically allowed $({\cal F},{\cal P})$ plane. Guided by the numerical exploration mentioned above, we have identified five different decompositions of $\hat{\cal U}$, depending on the region of the ${\cal F}$ vs. ${\cal P}$ distribution. These are provided in Fig.~\ref{gates} (although such figure refers to the specific case of $p_w=0.75$, the quantum circuit decompositions are independent of the initial value of the purity being considered). However, this leaves the parameters of the rotations involved in such circuits undetermined. In particular, we are interested in finding the values associated with the top-most boundaries in Fig.~\ref{gates}, {\it i.e.} the rotation angles $\{v=\alpha,\beta\}$ that allow us to achieve the largest state fidelity per set values $\tilde{\cal P}$ of the purity. This problem can be efficiently formulated in terms of the Lagrange-multiplier formalism. We have thus considered the functional
\begin{equation}
{\cal L}_{j}(v_j)={\cal F}_j(v_j)+\lambda_j[\tilde{\cal P}-{\cal P}_j(v_j)],
\end{equation}
where $j=A,..,E$ is a label identifying the trait j of the boundary considered each time, ${\cal F}_j(v_j)$ [${\cal P}_j(v_j)$] is the corresponding average fidelity [purity] and $\lambda_j$ is an unknown Lagrange multiplier. ${\cal L}$ is extremized by solving the set of simultaneous equations $\partial_{v_j}{\cal L}(v_j){=}0~~~\forall{v_j}$, with additional constraint ${\cal P}_j(v_j)=\tilde{\cal P}$. The results are shown as the solid curve enclosing the distribution of points in the $({\cal F},{\cal P})$ plane of Fig.~\ref{gates}. This completely solves the problem addressed in this work.

\noindent
{\bf Interpretation.} We now aim at understanding this result from an information theoretical viewpoint. In particular, we show that the impossibility to increase the output state fidelity past the point at ${\cal P}{=}{\cal P}_{in}$ could be related to the information cost of quantum state merging~\cite{merging,merging1}, a primitive that we now shortly describe. Consider two random variables $A$ and $S$, accessible to Aidan and Susy, respectively. In particular, information over $A$ completes the one brought about by $S$. How much information should Aidan send to Susy if she aims at having full information over his random variable? Slepian and Wolf provided the answer to this question, which is a fundamental point of classical information theory, finding that the amount of information required for the task is given exactly by the conditional entropy $H(A|S)=H(AS)-H(S)$ with $H$ the Shannon entropy~\cite{SW}. State merging~\cite{merging,merging1} is the quantum extension of the Slepian-Wolf result and considers Aidan and Susy as holding qubits $a$ and $s$, respectively, prepared in state $\varrho_{as}$. We now take a purification $\ket{\psi}_{ase}$ of such density matrix. Here, with no loss of generality, $e$ is a second qubit held by Susy, who also has available a `storing qubit' $e'$ whose role will be clarified soon. Which is the minimal quantum information that Aidan should send to Susy in order for her to construct a state $|{\tilde\psi}\rangle_{see'}$ close to $\ket{\psi}_{ase}$? The answer to such question (which clearly embodies a strict analogy of the scenario addressed by Slepian and Wolf) states that, if the two parties have $n\rightarrow\infty$ copies of the purification, with asymptotically vanishing errors, such amount of information is given by the quantum conditional entropy ${\cal E}(\varrho_a|\varrho_{s}){=}S(\varrho_{as})-S(\varrho_{s})$, where $S$ is the von Neumann entropy~\cite{merging,merging1}. The interesting part stays in the physical consequences related to the `sign' of ${\cal E}(\varrho_a|\varrho_{s})$: when negative, Susy can get the full state with LOCC operations only and is able to distill $-{\cal E}(\varrho_a|\varrho_{s})$ ebits of entanglement per copy of the purification that could be later used as a resource. Differently, a positive conditional entropy implies that state merging can be successful only when a sufficient amount of entanglement per copy of the purification is consumed.

\begin{figure}[b]
\psfig{figure=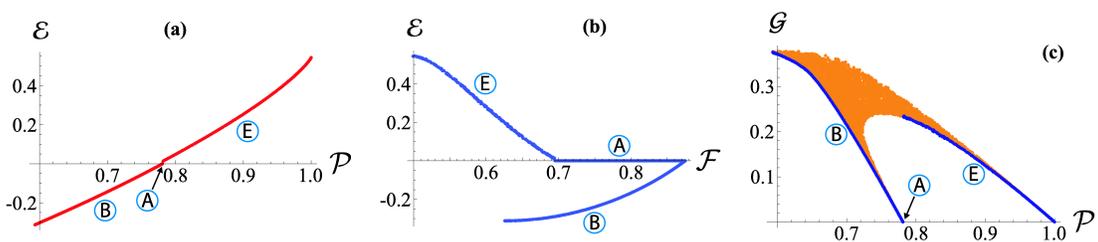,width=14.5cm}
\caption{{\bf (a)} Average conditional entropy ${\cal E}$ corresponding to the unitary operations $\hat{\cal U}$ lying along the boundaries in Fig.~\ref{gates} studied against the average output purity ${\cal P}$. {\bf (b)} Same as in panel {\bf (a)} but for the fidelity ${\cal F}$ being shown in the horizontal axis. The relevant boundary zones are clearly labelled as done in Fig.~\ref{gates}. Notice that we only show the parts of the boundary curve close to the zone of sign-flip of the conditional entropy. {\bf (c)} Geometric discord ${\cal G}$ against the average output purity ${\cal P}$ for ${\cal P}_{in}=0.78125$. The (orange) points show the average geometric discord obtained using a random-matrix sample analogous to the one used in Figs.~\ref{distribuzioni} and \ref{gates}, while the blue lines are the curves that correspond to part of the unitary operations $\hat{\cal U}$ lying along the boundaries in Fig.~\ref{gates}. Interestingly, these are also the unitary operations that minimize ${\cal G}$ at set values of output state purity. As before, we are only concerned with the region close to the sign-flip of the conditional entropy.}
\label{condent}
\end{figure}

We have thus calculated the conditional entropy associated with the states giving rise to the points along the boundary of the ${\cal F}$ vs. ${\cal P}$ graph in Fig.~\ref{gates}, and evaluated its average value in the same way as previously done for the average output fidelity and purity. We are interested in three different regions, close to ${\cal P}={\cal P}_{in}$. For ${\cal P}<{\cal P}_{in}$ and large values of the state fidelity (therefore corresponding to trait \circled{B}), the conditional entropy is negative and decreasing in modulus when the purity is increasing, as shown in Fig.~\ref{condent} {\bf (a)}. Reaching ${\cal P}={\cal P}_{in}$, the conditional entropy becomes null and so remains across the whole trait \circled{A} of the boundary curve, as shown in Fig.~\ref{condent} {\bf (b)}. This is the region of sign-flip of ${\cal E}$ that marks the change in resources needed in order to run the state merging protocol: for ${\cal P}>{\cal P}_{in}$, {\it i.e.} if the output state of our process is indeed purified and we want to retain a large state fidelity (upper part of trait \circled{E}), the conditional entropy is positive and increasing when the purity is increasing, as in the case of state merging requiring the consumption of a growing number of ebits. The connection between quantum state merging and the purification protocol addressed here allows us to explore even further interesting implications of our work. It has been found in Ref.~\cite{datta} that the minimum increase in the average cost of quantum state merging, when a measurement is performed on Aidan side, is exactly equal to the quantum correlations within $\varrho_{as}$ as measured by quantum discord~\cite{discord,discord1} (see also Ref.~\cite{cavalcanti} for a related result). The latter quantifies the genuine content of quantum correlations (beyond entanglement) shared by the parties in a bipartite state. Among the various formulations of measures for quantum correlations, geometric discord for two qubits can be easily calculated~\cite{borivoje,kavan} [cf. {\bf Methods} for the formal definition of discord and its geometric version]. Through the usual random-unitary approach, we have explored the distribution of the average output geometric discord against the average output purity for ${\cal P}_{in}=0.78125$ (qualitatively similar results are found for any other value of the input state purity). The results, shown in Fig.~\ref{condent} {\bf (c)}, are quite interesting:  at a set value of output purity, the amount of quantum correlations shared by system and ancilla is constrained to a non-convex region showing the existence of upper and lower bounds. By using again the Lagrange-multiplier approach, we have found that the upper part of trait \circled{E} corresponds to the curve that minimizes the degree of geometric discord at set output purity, for ${\cal P}>{\cal P}_{in}$. Quite intuitively, the points along trait \circled{A} accumulate on a single point at null discord and purity equal to ${\cal P}_{in}$. Approaching this point moving along \circled{E} for decreasing ${\cal P}$ results in an abrupt jump in the values of the output discord. On the other hand, when we move along \circled{B}, for increasing values of ${\cal P}$ towards ${\cal P}_{in}$, the geometric discord smoothly vanishes. Also in this case, trait \circled{B} corresponds to the curve that minimizes the degree of geometric discord at set output purity, for ${\cal P}<{\cal P}_{in}$. This clearly shows the intimate relation between the no-go result that is the key of our study and deep information theoretic concepts addressing information and quantum correlations. Needless to say, a qualitatively similar behaviour is found for any measure of bipartite entanglement (for instance, we have checked that analogous conclusions are reached by studying the distribution and bounding curves when negativity is used).
\\
\\
\noindent
{\large {\bf Discussion}}

\noindent
We have explored the trade-off between purity and fidelity in a state purification protocol, finding that such quantities are related in a highly non-trivial way, intimately connected to the nature, as a resource, of the ancilla-system state. Seen from a complementary perspective, our {\it no-go} result can also be interpreted as an attempt to quantify the amount of information that we can transfer to the ancilla $a$ by slowly decreasing its purity, until the amount that would be passed using a (classical) $\hat{\rm SWAP}$ gate (and local unitaries) is reached. It will be interesting to extend the breath of such findings, looking for similar negative results when addressing bipartite states and the distillation of entanglement, as well as seek for an experimental verification of the predictions of our analysis.\\
\\
\\
\noindent
{\large {\bf Methods}}\\
\noindent
{\bf Random Unitary Matrices.} Here we briefly outline the recipe to generate a random unitary matrix. The parameterization is based on the original work presented by Hurwitz in 1897~\cite{hurwitz}. Any unitary matrix, $U_r$, of dimension $n$, can be decomposed as
\begin{equation}
U_r=e^{i \alpha}E_1E_2\dots E_{n-1}
\end{equation}
where $E_{i}$ is an $n\times{n}$ matrix. Matrices $E_{i}$'s are readily constructed using products of proper rotation matrices $R^{(i,j)}(\phi_{i\,j},\psi_{i\,j},\chi_{i\,j})$, each depending on the respective set of Euler's angles $\{\phi_{i\,j},\psi_{i\,j},\chi_{i\,j}\}$ as follows
\begin{equation}
\begin{aligned}
E_1&=R^{(1,2)}(\phi_{1\,2},\psi_{1\,2},\chi_{1\,2}),\\
E_2&=R^{(2,3)}(\phi_{2\,3},\psi_{2\,3},0)R^{(1,3)}(\phi_{1\,3},\psi_{1\,3},\chi_{1\,3}),\\
E_3&=R^{(3,4)}(\phi_{3\,4},\psi_{3\,4},0)R^{(2,4)}(\phi_{2\,4},\psi_{2\,4},0),\\
&\times{R}^{(1,4)}(\phi_{1\,4},\psi_{1\,4},\chi_{1\,4}),\\
&\vdots\\
E_{n-1}&=R^{(n-1,n)}(\phi_{n-1\,n},\psi_{n-1\,n},0)\times\dots\\
&\times{R}^{(1,n)}(\phi_{1\,n},\psi_{1\,n},\chi_{1\,n}).
\end{aligned}
\end{equation}
The matrix elements are taken as
\begin{equation}
\begin{aligned}
R^{(i,j)}_{k,k}&=1~~(\text{for}~k\neq i,j),\\
R^{(i,j)}_{i,i}&=e^{i\psi}\cos\phi,~~R^{(i,j)}_{i,j}=e^{i\chi}\sin\phi,\\
R^{(i,j)}_{j,i}&=-e^{-i\chi}\sin\phi~~R^{(i,j)}_{j,j}=e^{-i\psi}\cos\phi,\\
\end{aligned}
\end{equation}
and zero otherwise. The angles are drawn from the ranges $\phi_{i\,j}\in[0,{\pi}/{2}],\psi_{i\,j}\in[0,2\pi],\chi_{i\,j}\in[0,2\pi]$ and $\alpha\in[0,{2\pi}]$, uniformly with respect to the corresponding Haar measure~\cite{kus1994}.\\
\noindent
{\bf Quantum Discord.} As originally proposed ~\cite{discord,discord1}, quantum discord can be associated with the difference between two classically equivalent versions of mutual information, which measures the total correlations within a quantum state. For a two-qubit state $\rho^{(r,r')}$, the mutual information is defined as ${\cal I}(\rho^{(r,r')})\,{=}\,{S}(\rho^{(r)}){+}{S}(\rho^{(r')}){-}{S}(\rho^{(r,r')})$. Alternatively, one can consider the one-way classical correlation ${\cal J}^\leftarrow(\rho^{(r,r')})\,{=}\,{ S}(\rho^{(r)}){-}{\cal H}_{\{\hat\Pi_i\}}(r|r')$\cite{discord,discord1}, where we have introduced ${\cal H}_{\{\hat\Pi_i\}}(r|r'){\equiv}\sum_{i}p_i{S}(\rho^i_{r|r'})$ as the quantum conditional entropy associated with the the post-measurement density matrix $\rho^i_{r|r'}\,{=}\,\text{Tr}_{r'}[\hat\Pi_i\rho^{(r,r')}]/p_i$ obtained upon performing the complete projective measurement $\{\Pi_i\}$ on spin $r'$ ($p_i\,{=}\,\text{Tr}[\hat\Pi_i\rho^{(r,r')}]$). Discord is thus defined as
\begin{equation}
{\cal D}^\leftarrow\,{=}\,\inf_{\{\Pi_i\}}[{\cal I}(\rho^{(r,r')}){-}{\cal J}^\leftarrow(\varrho^{(r,r')})]
\end{equation}
with the infimum calculated over the set of projectors $\{\hat\Pi_i\}$\cite{discord,discord1}. Analogously, one can define ${\cal D}^\rightarrow$, which is obtained upon swapping the roles of $r$ and $r'$.

Quantum correlations can also be defined by taking a geometric perspective and quantifying them as the minimum distance between a given state $\rho^{(r,r')}$ and the set of states $\sigma$ that are left unmodified by at least one measurement operated on one of the qubits. Therefore, by assuming the Hilbert-Schmidt norm $d(a)=\sqrt{\text{Tr}(aa^\dag)}$ as a metric (with $a$ an arbitrary square matrix), we can define the geometric discord as 
\begin{equation}
D_{\cal G}=2\min_{\sigma} d^2(\sigma-\rho^{(r,r')}),
\end{equation}
where the minimization is performed over all possible $\sigma$ defined above. The explicit calculation of $D_{\cal G}$ is possible without heavy computational efforts and, actually, analytically as illustrated in Refs.~\cite{borivoje,girolami}. 

\vspace{.5cm}
\noindent
{\large {\bf Acknowledgments}}\\
\noindent
We thank F. Fanchini, X. Ma, T. Paterek, and F. Sciarrino for discussions. This work was supported by  the UK EPSRC under a Career Acceleration Fellowship and a grant of the ``New Directions for Research Leaders" initiative (EP/G004579/1). 

\end{document}